\documentclass[preprint,prd,aps,showpacs,showkeys,onecolumn,english]{revtex4}
\usepackage{amsmath,amssymb} 
\usepackage[dvips]{graphicx} 
\usepackage{graphics}
\usepackage{graphicx}
\usepackage{epsfig}
\usepackage{slashed}
\usepackage[english]{babel}
\usepackage{amsmath}
\usepackage{amssymb}
\usepackage{graphics}
\usepackage{graphicx}
\usepackage{epsfig}
\usepackage{slashed}
\usepackage{subfigure}	
\usepackage[export]{adjustbox}

\newcommand\nn{\nonumber}
\newcommand\ba{\begin{eqnarray}}
\newcommand\ea{\end{eqnarray}}
\newcommand{\br}[1]{\left( #1 \right)}

\newcommand{\GeV}{~\mbox{GeV}}
\newcommand{\MeV}{~\mbox{MeV}}

\newcommand{\eV} {~\mbox{eV}}
\newcommand{\Sp}{\mbox{Sp}}

\begin{document}
\title{Total cross section of the process $e^+ + e^- \to \Sigma^0 + \bar{\Sigma}^0$
in the vicinity of charmonium $\psi (3770)$ including the $D$-meson loop and three gluon contributions}
\author{Azad I.~Ahmadov$^{a,b}$ \footnote{E-mail: ahmadov@theor.jinr.ru}}
\affiliation{$^{a}$ Institute of Physics,
Ministry of Science and Education, H.Javid ave., 131, AZ1143 Baku, Azerbaijan}
\affiliation{$^{b}$ Bogoliubov Laboratory of Theoretical Physics,
JINR, Dubna, 141980 Russia}

\date{\today}

\begin{abstract}
For the study of the structure of baryons it is necessary to investigate the production
of a baryon pair in $e^+e^-$ annihilation. The baryon-antibaryon pair production at the electron-positron
linear collider makes it possible to investigate in detail the basic structure of the Standard Model.
The creation of baryon-antibaryon pairs in electron-positron annihilation provides an
increasingly powerful tool at higher c.m. energies.
We present phenomenological results for $\Sigma^0 \bar {\Sigma}^0$ production in
$e^+e^-$ interaction at the BESIII and BABAR Colliders.
In the present work, we investigate a hyperon pair produced in the reaction
$e^+e^- \to \Sigma^0 \bar{\Sigma}^0$.
We calculate the total cross section of the process $e^+e^- \to \Sigma^0 \bar {\Sigma}^0$ taking into account the
contributions of the $D$-meson loop and three gluon loops as well as the interference of all diagrams to the
Born approximation.
For these contributions large relative phases are generated with respect to the
pure electromagnetic mechanism.
For the large momentum transferred region we obtain as a byproduct a fit of the electromagnetic
form factor of the $\Sigma$ hyperon.
\vspace*{0.5cm}

\noindent
\end{abstract}

\maketitle
\section{Introduction}
\label{Introduction}

It is understandable that the Universe consists of baryons, the lightest baryons and
nucleons account essentially for all of the observable matter.
In fact, understanding the structure of nucleons is of high importance
for the whole set of baryons forming the lowest SU(3) octet.
To understand the strong interaction in the confinement domain, i.e.,
where quarks form hadrons is one of the most challenging questions in
contemporary physics.

The study of the electromagnetic structure of hadrons, which began with
the pioneering work of Hofstadter \cite{I1,I2}, up to now remains an open and
interesting field of research in high energy physics.
We can say that one of the most basic tools for studying the structure
of the nucleon is the production of hadrons in electron-positron interaction
at high energies.

The electromagnetic form factors (FFs), which can be investigated using
the process of electron-positron interaction, are among the most basic
quantities containing information about the internal structure of the nucleon.
The electromagnetic form factors of nucleons with a large momentum
transfer provide valuable information about their structure.
In \cite{Haidenbauer}, the cross section of the reactions
$\bar {p} p \to e^+e^-$ and $e^+e^- \to \bar {p} p$ are analyzed in the near-threshold region and
are used for the proton of the effective form factor
(form factors $G_E$ and $G_M$) at energies close to the $\bar {p} p$ threshold.
The authors of \cite{Bianconi} highlighted the presence in the BABAR
data of deviations in the timelike form factors of protons from the
point behavior of the proton-antiproton electromagnetic current
in the reaction $e^+e^- \to \bar {p} p$.
Therefore, in this work, a form factor was used in the form of $F_0$ + $F_{osc}$,
where $F_0$ is a parametrization expressing the long-term trend of the
form factor, and $F_{osc}$ is a function of the form exp(-Bp)cos(Cp),
where $p$ is the relative momentum of the final $\bar {p} p$-pair.
In \cite{Qin}, the effective EMFFs of the proton and neutron
in the timelike region are investigated at the electron-positron
annihilation into antinucleon-nucleon ($\bar {N} N$) pairs, which are treated
in the distorted wave Born approximation.
In \cite{Milstein}, the method of effective optical
potential that well describes the $N \bar {N}$ scattering phases and a sharp
dependence of the $N \bar {N}$ production cross sections was used for the
$p \bar {p}$ and $n \bar {n}$ production in e+e annihilation
near the threshold.
The authors of this work used the electromagnetic
form factors $G_E$ and $G_M$ for protons and neutrons near the threshold.

In \cite{I3,I4} it was proposed that the electromagnetic form factors of hadrons can be studied
for timelike region momentum transfers, $q^2>0$, through measuring hadron pair production cross sections
in electron-positron collisions process.

The electromagnetic form factor in the timelike region  in terms of available
data \cite{I5,I6,I7,I8,I9,I10,I11,I12,I13} is consistent with naive quark counting rules
and the perturbative QCD (pQCD) prediction at large $q^2$  \cite{I14,I15}.
In this case, the use of the timelike form factors can offer a unique
opportunity  to study the inner structure of hadrons and the electromagnetic
property of hyperons.

Today, most of our knowledge of the experimentally established nucleon
resonances listed in \cite{I16} is mainly from the electron-positron and proton-proton
(proton-antiproton) interaction experiments.

It is necessary to note that in the transition region between the perturbative
and nonperturbative regimes the $\psi (3770)$ has a mass
of the charmonium resonance.
Therefore, to study the structure of baryons one should investigate
the production of a baryon pair in $e^+e^-$ annihilation.
Therefore, the study of strong and hadronic decays of the $\psi (3770)$ will
provide knowledge of its structure in perturbative and
nonperturbative strong interactions in this energy region \cite{I17}.

It must be emphasized that the BESIII have collected the largest data sample of
e+e- collisions at 3.773 GeV.
After analyzing samples together with data, it became possible to
theoretically investigate exclusive decays of $\psi(3770)$
taking into account the interference of resonant and nonresonant amplitudes.

The charmonium states with $J^{PC}=1^{--}$, such as $J/\psi$,  $\psi(3770)$,
and others, are productions through electron-positron annihilation into
a virtual photon at electron-positron colliders.
This is followed by a decay in  these charmonium states, i.e.,
decay into light hadrons through either the three-gluon process
($e^+e^- \to \psi \to ggg \to hadrons$), or the one-photon process ($e^+e^- \to \psi \to \gamma^*  \to hadrons$).

According to the Okubo-Zweig-Iizuka (OZI) rule, the $\psi(3770)$, the lowest
lying $1^{--}$ charmonium state above the $D \bar{D}$ threshold, is expected to
decay dominantly into the $D \bar{D}$ final states \cite{I18,I19}.
In \cite{I20}, it was shown that the binary with two particles produced in the final
states give the possibility to further simplify the consideration of the
processes with charmonium in the intermediate state.

The study of the process of $\psi(3770)$ production in e+e- annihilation and
its subsequent decay into two hadron is a test of the prediction of QCD,
which can be understood based on quark distribution amplitudes
in hadron-hadron pairs, and the total hadron helicity conservation.
Due to its richness of $c\bar{c}$ states the $\psi(3770)$ is one of those prominent
structures in the hadronic cross section, that are of great interest to theory \cite{I16}.

At the energy around 3.770 GeV, the well-established $\psi(3770)$ resonance is the only observed
structure, i.e., the $\psi(3770)$ is the lowest mass charmonium resonance above the open charm
pair $D\bar{D}$ production threshold.
It is expected that the $\psi(3770)$ resonance can decay almost entirely into a
pure $D\bar{D}$ \cite{I21} and the baryon-antibaryon pair production at an electron-positron
collider can be tested by fundamental symmetries in the baryon sector,
in particular when the probability of the process is enhanced by a resonance
such as the $J/\psi$ (or $\psi$) \cite{I22}.

It is necessary to note that the BESIII Collaboration performed high-precision studies of
a possible threshold enhancement in the
$e^+e^- \to \Sigma^{\pm} \bar{\Sigma}^{\mp}$ \cite{I23} and
$\Xi^-\bar{\Xi}^+$ \cite{I24} processes and also showed that the cross section
is nonvanishing near the threshold.
This means that the threshold effect obtained in this way will be useful for
measuring the near-threshold pair generation of hyperons
$\Sigma^0\bar{\Sigma}^0$, which was observed in the BESIII\cite{I37} and BABAR\cite{I9} experiments.

We should note that the $e^+e^- \to \Sigma^ 0 \bar {\Sigma}^0$ process has been
studied in detail by many authors, and several experiments are extremely
important for modern high-energy physics \cite{I9,I25,I26, I27,I28,I29,I30,I31,I32,I33,I34,I35,I36,I37}.

In the present paper, the first problem is the calculation of the total cross section
with allowance for the $D$-meson loop and three gluon contributions
within the framework of QED and the SM.

Our aim in this paper is to study the characteristics of the
$\Sigma^0 \bar{\Sigma}^0$ production process with taking into account the
contributions of the $D$-meson loop and three gluon loops in the Born approximation.

\section{The process $e^+e^- \to \Sigma^0 \bar{\Sigma}^0$ in Born approximation \label{Born}}

In this section, in the leading order we want to consider the $\Sigma^0 \bar{\Sigma}^0$ production in the process electron-positron annihilation.
The process is written in the form,
\ba
e^+(p_1) + e^-(p_2) \to \Sigma^0(q_1) + \bar{\Sigma}^0(q_2),
\label{A1}
\ea
where $p_1, \, p_2$ are 4-momenta of the positron and electron in the initial state; $q_1, q_2$ are 4-momenta of the
$\Sigma^0$ and $\bar{\Sigma}^0$ in the final state.

The kinematics for the process can be written in terms of the following Mandelstam invariants:
\ba
s &=&(p_{1} + p_{2})^{2} = (q_{1} + q_{2})^{2};  \notag \\
t &=&(p_{1} - q_{1})^{2} = (p_{2} - q_{2})^{2};  \notag \\
u &=&(p_{1} - q_{2})^{2} = (p_{2} - q_{1})^{2},
\label{Mand1}
\ea
In this process, it follows from \eqref{Mand1} that the sum of Mandelstam invariants, which we will use, can be connected,  in the form
\ba
s + t + u = 2m_{e}^{2} + 2 M_{\Sigma}^2,
\label{3}
\ea
where $m_{e}$ and $M_{\Sigma}$ are the electron and $\Sigma^0$ hyperon masses, respectively.
In this work, we neglect the mass of the electron $m_e$.

\begin{figure}
    \centering
    \includegraphics[width=0.40\textwidth]{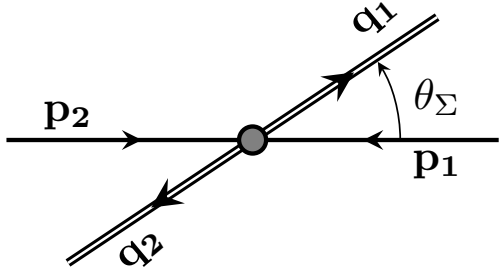}
    \caption{The definition of the scattering angle $\theta_\Sigma$ from (\ref{Phi2}) in
    the c.m.system.}
    \label{fig.MomentaPosition}
\end{figure}

The "master formula" for evaluating  the cross section for the process \eqref{A1} has the form:
\ba
d\sigma = \frac{1}{8 s} \sum_{\text{spins}} |{M}|^2 \, d\Phi_2,
\label{CrossSectionGeneralForm}
\ea
the square of the matrix element is summed over all possible spin states of the initial and final particles. \\
The phase-space element of final particles $d\Phi_2$  can be written in the following form:
\ba
&& d\Phi_2 = (2\pi)^4 \delta (p_1 + p_2 - q_1 - q_2) \frac{d\bf{q_1}}{(2\pi)^3 2E_1}\frac{d\bf{q_2}}{(2\pi)^3 2E_2} = \nn\\
&&= \frac{1}{(2\pi)^2} \delta (p_1 + p_2 - q_1 - q_2) \frac{d\bf{p_1}}{2E_1}\frac{d\bf{p_2}}{2E_2} =
    \frac{|\bf{p}|}{16 \pi^2 \sqrt{s}} \, d\Omega_\Sigma =\frac{\beta}{16 \pi} \, d\cos\theta_\Sigma,
    \label{Phi2}
\ea
where $d\Omega_\Sigma=d\phi_\Sigma \, d\cos\theta_\Sigma = 2\pi\, d\cos\theta_\Sigma$, and
$\phi_\Sigma$ and $\theta_\Sigma$ are the azimuthal and the polar angles of the final $\Sigma$-hyperon momentum
in the c.m.system, that is $\theta_\Sigma$ is the angle between the directions of the momenta
of the initial electron $\bf{p_2}$ and the final $\Sigma$-hyperon $\bf {q}_1$ (Fig.~\ref{fig.MomentaPosition}).
The modulus of three momenta of the final $\Sigma^0$-hyperon (or $\bar {\Sigma}^0$-hyperon) is fixed in this case
by the $\delta$-function in the phase volume, i.e.,
\ba
&& |{\bf{p}}| \equiv |{\bf{q_1}}| = |{\bf{q_2}}| = \frac{\sqrt{s}}{2}\sqrt {1 - \frac{4M^2_{\Sigma}}{s}} = \frac{\sqrt{s}}{2} \beta, \\ \nn
&& \beta = \sqrt {1 - \frac{4M^2_{\Sigma}}{s}},
\ea
where $\beta$ is the velocity of the $\Sigma^0$-hyperon in the $e^+e^-$ c.m.system, $s$ is the square of the
c.m.energy, and $M_{\Sigma}$ is the mass of the $\Sigma^0$-hyperon.

\begin{figure}
	\centering
    \subfigure[]{\includegraphics[width=0.40\textwidth]{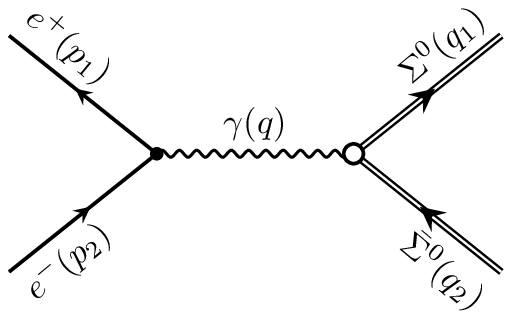}\label{BornDiagram}}
	\hspace{0.05\textwidth}
    \subfigure[]{\includegraphics[width=0.40\textwidth]{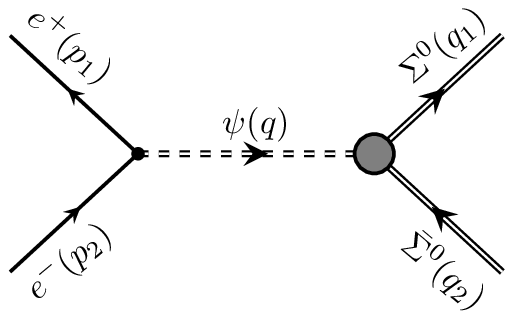}\label{PsiDiagram}}
    \caption{The Feynman diagrams describing for the $e^+e^- \to \Sigma^0 \bar{\Sigma}^0$ process corresponding
            to the Born approximation (a) and the intermediate state $\psi(3770)$ charmonium (b).}
    \label{fig.TwoMechanisms}
\end{figure}

The four-momenta of the leptons and $\Sigma^0$-hyperons are given by
\ba
p_2 = \frac{\sqrt{s}}{2}(1,0,0,1),\,\,\,\,p_1 = \frac{\sqrt{s}}{2}(1,0,0,-1),  \nn \\
q_1 = \frac{\sqrt{s}}{2}(1,\beta \sin\theta_\Sigma,0, \beta \cos\theta_\Sigma), \nn
q_2 = \frac{\sqrt{s}}{2}(1,-\beta \sin\theta_\Sigma,0,-\beta \cos\theta_\Sigma). \nn
\ea
The Feynman diagram for $e^+e^-$ annihilation into a virtual photon, with further production of the $\Sigma^0 \bar{\Sigma}^0$
pairs in the Born approximation of the process \eqref{A1} is illustrated in Fig.~\ref{BornDiagram}.

In the Born approximation, the process \eqref{A1} is described by a cleanly electrodynamic diagram with Fig.~\ref{BornDiagram}.
We can write the matrix element corresponding to the Feynman diagrams for the process $e^+(p_1) + e^-(p_2) \to \Sigma^0(q_1) + \bar{\Sigma}^0(q_2)$
in the Born approximation as follows:
\ba
   {\mathcal M}_B = -\frac{e^2}{s} [\bar{v}(p_1) \gamma_\mu u(p_2)] \, [\bar{u}(q_1) \Gamma_\mu(q) v(q_2)],
    \label{BornAmplitude}
\ea
where the quantity $e$ is the elementary electric charge, i.e., $e=\sqrt {4\pi\alpha}$, $\alpha \approx 1/137$
is the fine structure constant \cite{I16}, and $s$ is the squared the total invariant mass of the lepton pair. \\
To describe the composite nature of the hyperons (baryons), form factors have been introduced.
As shown in Fig.~\ref{BornDiagram}, for the annihilation process, the $\gamma \Sigma^0 \bar {\Sigma}^0$ $(\gamma B \bar {B})$
current can be written in terms of the Pauli Form Factors $F_1$ and $F_2$.
The vertex function $\Gamma_\mu(q)$, which describes the vertex of the photon with hyperons [Fig.~\ref{BornDiagram}] can be written as follows:
\ba
	\Gamma_\mu(q) &= F_1(q^2) \gamma_\mu - \frac{F_2(q^2)}{4M_\Sigma} (\gamma_\mu \hat{q} - \hat{q} \gamma_\mu),
\label{FF}
\ea
where $M_{\Sigma}$ is the hyperon (baryon) mass, $q$ is the transferred momentum. Here, for $\hat{q}$ we will use the notation
$\hat{q} = q_\nu \gamma^\nu$.
In \eqref{FF} the functions $F_1(q^2)$ and $F_2(q^2)$ are the form factors of $\Sigma$-hyperons and are usually normalized as follows,
$F_1(0)=0$ and $F_2(0)=\mu_{\Sigma}$, where $\mu_\Sigma$ is the $\Sigma$-hyperon anomalous magnetic moment.

However, in the work \cite{I38} the authors showed that this pointlike behavior of the proton near the threshold is not so unambiguous,
i.e., the nontrivial structure of the baryon starts to manifest itself even at relatively low $q^2$ and thus one must
take into account the structure of these effects.
Moreover, a comparison of the Born cross section with the data below will show that we cannot do without a form factor.
However, it should be noted here that in the timelike region, due to small statistics and, consequently, due to large errors
in experimental data, it is not possible to distinguish between the electric $G_E$  and magnetic $G_M$ form factors in the experiment.

Therefore, we, like many authors, will use the approximation $|G_E| = |G_M|$, from which it follows that $F_2(q^2)=0$.
Thus, we introduce the effective form factor $F_1(q^2) = G(q^2)$, which was used on the basis of QCD in \cite{I14,I40}.

For the square of the matrix element by \eqref{BornAmplitude}, after calculating the trace and taking into
account the form factor $F_1(q^2) = G(q^2)$ the following formula can be obtained:
\ba
    \sum_{spins} |{\mathcal M}_B|^2
    &=
    64 \pi^2 \alpha^2 |{G(s)}|^2 ( 2 - \beta^2 \sin^2\theta_\Sigma).
    \label{AmplitudeSquare}
\ea
Using \eqref{AmplitudeSquare} and taking into account the expressions for the phase volume in \eqref{Phi2}, the differential
cross section \eqref{CrossSectionGeneralForm} can be written in the following form:
\ba
    d\sigma_B(s)=\frac{\pi\alpha^2 \beta}{2s}|{G(s)}|^2 (2-\beta^2 \sin^2\theta_\Sigma)\,\,d\cos\theta_\Sigma.
    \label{DTotalCrossSectionBorn}
\ea
To obtain the total cross section, it is necessary to integrate this expression \eqref{DTotalCrossSectionBorn} over all possible scattering angles
$d\cos\theta_\Sigma = \sin\theta_\Sigma \,d \theta_\Sigma.$
The integration limits for the angles are determined as follows:
\ba
0 \leq \theta_\Sigma \leq \pi.  \nn
\label{theta}
\ea
After integrating over the angle $\theta_\Sigma$, we obtain an expression for the total cross section in the Born approximation in the following form:
\ba
    \sigma_B(s)=\frac{2\pi\alpha^2}{3s} \beta (3-\beta^2) |G(s)|^2.
    \label{TotalCrossSectionBorn}
\ea
The form factor $G(s)$, for which we use the pQCD form from \cite{I14,I40}, which takes into account the running of the QCD coupling constant $\alpha_s$:
\ba
	G(s) = \frac{C}{s^2 \log^2\br{s/\Lambda_{QCD}^2}},
	\label{Formfactor}
\ea
where $C$ is a free parameter, $\Lambda_{QCD}$ is the QCD scale parameter.
It is necessary to note that the constant $C$ should be fitted to the experimental data for hyperon-antihyperon production
in the energy range of the corresponding experiment.

\begin{figure}
	\centering
	\includegraphics[width=0.55\textwidth]{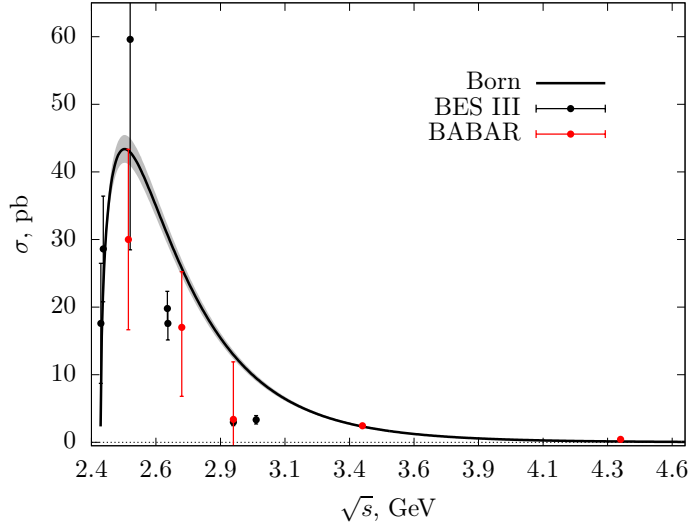}
    \caption{The total cross section for the process $e^+ e^-\to \Sigma^0 \bar{\Sigma}^0$ as a
    function of the c.m.energy.
    Black line is the total cross section in Born approximation (\ref{TotalCrossSectionBorn}).
    The curve errors origin from the form factor constant (\ref{eq.C}) fitting errors.}
    \label{WideRange}
\end{figure}

In the present work of $\Sigma^0\bar{\Sigma}^0$ pair production, we fix this constant using the BES-III measurement \cite{I37} presented in
Fig.~\ref{WideRange}.
At the value $\Lambda$ = 300 \,MeV for the parameter $C$ after fitting in the Born cross section from (\ref{TotalCrossSectionBorn}) with
respect to these data we have the value,
\ba
	C = (59.12 \pm 1.6)~\GeV^4,
    \label{eq.C}
\ea
which we will use further for the $\Sigma$-hyperon electromagnetic form factor (\ref{Formfactor}).
It should be noted that the equation for the form factor $G(s)$ (\ref{Formfactor}) with
the constant $C$ from (\ref{eq.C}) works for a relatively large momentum transferred $q^2$.
Here, it does not pretend to work near the threshold, since the Coulomb-like enhancement factor
with many delicate features plays an important role \cite{Haidenbauer,Amoroso}, or a manifestation of the wavelike nature of
hyperon (baryon)  of the stabilization after its emerging from the vacuum \cite{Egle}.

%
\section{The quarkonium $\psi(3770)$ intermediate state}
\label{sec.PsiIntermediateState}

We would like to note that in the process \eqref{A1} the main task in this work is to describe the effect of excitation of the
$\psi(3770)$ charmonium resonance.
As seen in Fig.~\ref{WideRange}, the total cross section in the Born approximation \eqref{A1}, which includes only the
electromagnetic mechanism, cannot describe this delicate behavior near the charmonium resonance $\psi(3770)$.
Therefore, in this region, it is necessary to take into account the additional contribution to the amplitude that appears
from the diagram with $\psi(3770)$ in the intermediate state (Fig.~\ref{PsiDiagram}) and is enhanced by the Breit-Wigner propagator.
Thus, we can calculate the contribution of the additional mechanism, when the $\psi(3770)$ ($I^G (J^{PC})=0^-(1^{--})$) charmonium
resonance is excited in the intermediate state.
In this case, the total amplitude of the process \eqref{A1} will be the sum of two amplitudes,
\ba
\mathcal {M} = \mathcal {M}_B + \mathcal {M}_{\psi},
\label{M}
\ea
where $\mathcal {M}_B$ is the amplitude \eqref{BornAmplitude} of the process \eqref{A1} in the Born approximation (Fig.~\ref{BornDiagram}),
and $\mathcal {M}_{\psi}$ already takes into account the contribution of the intermediate state of $\psi(3770)$ charmonium, the enhanced
by the Breit-Wigner factor [Fig.~\ref{PsiDiagram}],
\ba
    M_\psi = \frac{1}{s-M_\psi^2+i M_\psi\,\Gamma_\psi} J^{e\bar{e}\to\psi}_\mu(q) \biggl(g^{\mu\nu} - \frac{q^\mu q^\nu}{M_\psi^2}\biggr)
    J^{\psi\to \Sigma^0\bar{\Sigma}^0}_\nu (q),
    \label{Mpsi}
\ea
where $M_{\psi}$ = 3773.7 \,MeV and $\Gamma_{\psi}$ = 27.2 \,MeV \cite{I16} are the mass and total decay width of $\psi (3770)$ resonance;
$J^{e\bar{e}\to\psi}_\mu(q)$ and $J^{\psi\to \Sigma^0\bar{\Sigma}^0}_\nu (q)$ are the currents that describe the transition of an electron-positron
pair into the $\psi(3770)$ resonance and the transition of the $\psi(3770)$ resonance into a $\Sigma^0\bar{\Sigma}^0$ pair, respectively.
We take into account the fact that the currents $J^{e\bar{e}\to\psi}_\mu(q)$ and $J^{\psi\to \Sigma^0\bar{\Sigma}^0}_\nu (q)$ in \eqref{Mpsi}
have to be conserved, that is $q^{\mu}J^{e\bar{e}\to\psi}_\mu(q) = q^{\mu}J^{\psi\to \Sigma^0\bar{\Sigma}^0}_\mu (q) = 0.$
Here, following the paper \cite{Ahmadov}, we assume that the vector current $J^{e\bar{e}\to\psi}_\mu(q)$ will have the same structure as in the case of
a photon according to \eqref{BornAmplitude}, i.e.,
\ba
    J^{e\bar{e}\to\psi}_\mu(q) = g_e \, [\bar{v}(q_+) \gamma_\mu u(q_-)],
    \label{Jeepsi}
\ea
only with another constant $g_e = F_1^{\psi\to \Sigma^0 \bar{\Sigma}^0}(M_\psi^2)$, which is equal to the value of the form factor
$F_1^{\psi\to \Sigma^0 \bar{\Sigma}^0}(M_\psi^2)$ on the mass of charmonium, i.e., is the value of the form factor of the vertex
$\psi \to \Sigma^0\bar{\Sigma}^0$ at the $\psi(3770)$ mass-shell [here we accept the same approximation as in the Born case and assume that
$F_2^{\psi\to \Sigma^0 \bar{\Sigma}^0}(M_\psi^2] = 0]$.
Knowing the total decay width of $\psi\to e^+ e^-$, which is equal to $\Gamma_{\psi\to e^+e^-} = 261~\eV$ \cite{I16}, one can calculate
this constant $g_e$,
\ba
    g_e= \sqrt{\frac{12\pi\Gamma_{\psi\to e^+e^-}}{M_\psi}} = 1.6 \cdot 10^{-3}.
\ea
It is shown in \cite{Kuraev} that the possible imaginary part of the vertex $e\bar{e} \to \psi$ is small, less than 10~\% of the real part.
Therefore, we neglect such a possible imaginary part.

We can now calculate the contribution of the intermediate charmonium to the cross section.
If we substitute the total amplitude from \eqref{M} in the general formula for the cross section \eqref{CrossSectionGeneralForm},
then for the total cross section, we obtain the following expression:
\ba
	&\sigma \sim |\mathcal {M}|^2 = ||\mathcal {M}_B| + e^{i\phi} |\mathcal {M}_{\psi}||^2
	=\nn\\
	&\qquad = |\mathcal {M}_B|^2 + 2\cos\phi |\mathcal {M}_B| \cdot |\mathcal {M}_{\psi}| + |\mathcal {M}_{\psi}|^2
	\sim \sigma_B + \sigma_{int} + \sigma_\psi,
	\label{TotalCrossSection}
\ea
where $\phi$ is the relative phase between the Born contribution $\mathcal {M_B}$ and the additional contribution $\mathcal {M}_\psi$. \\
Thus, we need to calculate only the contribution of charmonium $\sigma_\psi$ to the cross section and the contribution of the
interference of charmonium with Born $\sigma_{int}$.
Evaluating $\sigma_\psi$ and taking into account the Born cross section $\sigma_B$ from \eqref{TotalCrossSectionBorn} and
the interference contribution $\sigma_{int}$ with the phase $\phi$, one can calculate the total cross section including both
contributions using \eqref{TotalCrossSection} in the following form:
\ba
	\sigma_\psi = \biggl(\frac{\sigma_{int}}{2 \cos\phi \, \sqrt{\sigma_B}} \biggr)^2.
\label{sigmapsi}
\ea
Thus, we need to calculate the interference of the Born $\mathcal {M_B}$ contribution with the contribution that takes into
account charmonium in the intermediate state of the $\mathcal {M}_\psi$.
According to the general formula \eqref{CrossSectionGeneralForm}, the interference contribution to the cross section of the process
can be written in the standard form,
\ba
    d\sigma_{int} = \frac{1}{8s} \sum_{\text{spins}} 2\,\mbox{Re}[\mathcal {M}_B^+ \mathcal {M}_\psi] \, d\Phi_2,
    \label{dsigmaint}
\ea
In order to obtain the contribution to the total cross section, we need to integrate in \eqref{dsigmaint} over the phase space
of the final particles,
\ba
    \sigma_{int}(s)
    &=
    \frac{1}{4 s^2} \mbox{Re}\biggr\{
    	\frac{\sum_{s}
    	(J^{e\bar{e}\to\gamma}_\mu)^* J^{e\bar{e}\to\psi}_\nu}
    	{s-M_\psi^2+i M_\psi\,\Gamma_\psi} \cdot
    	\sum_{s'}
    	\int d\Phi_2
    	(J_{\gamma\to \Sigma^0\bar{\Sigma}^0}^\mu)^* J_{\psi\to \Sigma^0\bar{\Sigma}^0}^\nu \biggr\},
    \label{sigmaint}
\ea
where $\sum_s$ and $\sum_{s'}$ are the summations over the spin states of the initial and final particles, respectively.
We now use the method of invariant integration over the total final phase volume, the second term in \eqref{sigmaint}
can be written in the following form:
\ba
	&\sum_{s'} \int d\Phi_2	(J_{\gamma\to \Sigma^0 \bar{\Sigma}^0}^\mu)^* J_{\psi\to \Sigma^0 \bar{\Sigma}^0}^\nu
	= \frac{1}{3} \biggl( g^{\mu\nu} - \frac{q^\mu q^\nu}{q^2} \biggr)
	\sum_{s'} \int d\Phi_2 	\br{J_{\gamma\to \Sigma^0 \bar{\Sigma}^0}^\alpha}^* J^{\psi \to \Sigma^0 \bar{\Sigma}^0}_\alpha.
\label{JJ}
\ea
Using the explicit form containing lepton currents $J^{e\bar{e} \to \gamma}_\mu$ from \eqref{BornAmplitude} and $J^{e\bar{e}\to\psi}_\nu$ from \eqref{Jeepsi},
and applying the conservation of the currents, we can calculate
\ba
&&	\sum_{s}(J^{e\bar{e} \to \gamma}_\mu)^* J_{e\bar{e} \to \psi}^\mu
 = -e g_e \sum_{s} [\bar{u}(p_2) \gamma_\mu(q) v(p_1)] [\bar{v}(p_1) \gamma^{\mu} u(p_2)] \approx  \nn \\
&&  \approx
	-e g_e \, \Sp[\hat{p_2} \gamma_\mu \hat{p_1} \gamma^\mu] \approx 4 \, e \, g_e s.
\label{JJe}
\ea
Substituting this result into \eqref{sigmaint} and using the explicit form of the two-particle phase volume of the final
particles from \eqref{Phi2} and the invariant integration  method from \eqref{JJ}, we obtain the interference contribution to the
total cross section in the following expression in a simplified form:
\ba
    \sigma_{int}(s)
    = \frac{e g_e \beta}{48 \pi s}
       \mbox{Re}\biggl\{\frac{1}{s-M_\psi^2+i M_\psi\, \Gamma_\psi}
    	\int\limits_{-1}^1 d\cos\theta_{\Sigma^0} \sum_{s'}
    	(J_{\gamma\to \Sigma^0 \bar{\Sigma}^0}^\alpha)^* J^{\psi \to \Sigma^0 \bar{\Sigma}^0}_\alpha
    \biggr\}.
    \label{TotalCrossSectionInterference}
\ea
In formula \eqref{TotalCrossSectionInterference}, the subintegral expression describes all the dynamics of the transformation of charmonium
into the $\Sigma^0\bar{\Sigma}^0$ pair and can be expressed in a separate form,
\ba
	S_i(s)
    = \frac{e g_e \beta}{48 \pi s}
    	\int\limits_{-1}^1 d\cos\theta_{\Sigma^0} \sum_{s'}
    	(J_{\gamma\to \Sigma^0\bar{\Sigma}^0}^\alpha)^* J^{\psi\to \Sigma^0\bar{\Sigma}^0}_\alpha.
    \label{Si}
\ea
Thus, the interference contribution to the total cross section \eqref{TotalCrossSectionInterference} can be written  in the form:
\ba
	\sigma_{int}(s)	=
	\mbox{Re}\biggl(\frac{S_i(s)}{s-M_\psi^2+i M_\psi\, \Gamma_\psi}\biggr),
	\label{SigmaIntViaSi}
\ea
The subscript index $i$ in \eqref{Si} denotes different possible mechanisms of this transition.
For example, the usual OZI-permitted mechanism through the $D$-meson loop, shown in Fig.~\ref{fig.DD}, is quite possible.
Since the mass of the charmonium $\psi (3770)$ is above the threshold for the production of the $D\bar {D}$-pair,
it is natural to expect that the $D$-meson loop will be the main mechanism in this reaction.
However, since the excess over the thresholds is minor ($M_{\psi} -2 M_D \approx$ 39 MeV, i.e., relative to the characteristic
energies in the problem, this value is about 1\%),
we expect this contribution to be small and it is also necessary to consider other possible mechanisms.
We think that a significant contribution will be made by the OZI-forbidden mechanism shown in Fig.~\ref{fig.3G},
which occurs due to the three-gluon annihilation of charmonium into a $\Sigma^0\bar{\Sigma}^0$ pair.

Using the procedure, which is described in \cite{Bystritskiy} [Eqs.~(15) and (16)], and with the help of the interference
contribution (\ref{SigmaIntViaSi}) with the total relative phase between the Born amplitude $\mathcal {M}_B$ and the charmonium
contribution $\mathcal {M}_\psi$, we can reconstruct the total cross section.

\section{The \textit{D}-meson loop mechanism}
\label{sec.DMesonLoopMechanism}
\begin{figure}
	\centering
    \includegraphics[width=0.50\textwidth]{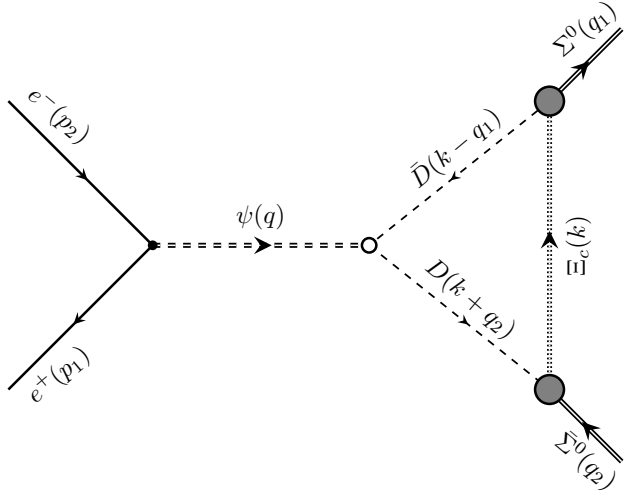}
    \caption{$D$-meson loop mechanism of the Feynman diagram contributing to the $\Sigma^0\bar{\Sigma}^0$ production in the process
    $e^+e^- \to \Sigma^0\bar{\Sigma}^0$ at the one-loop level.}
    \label{fig.DD}
\end{figure}

In this section, we will calculate the contribution of the intermediate charmonium with the transition to the final state
$\Sigma^0 \bar{\Sigma}^0$ hyperon pair through the $D$-meson loop, which is presented in Fig.~\ref{fig.DD}.
In order to calculate the contribution of the $D$-meson loop to the cross section, we need to calculate the quantity of $S_D$
from \eqref{Si}, which enters into \eqref{SigmaIntViaSi}.
For this, we have to construct the amplitude of the $\mathcal {M}_D$ corresponding to Fig.~\ref{fig.DD}, and then extract the
value of the $S_D$ from it.
Now we can write, according to Feynman's rules, the $D$-meson loop contribution to the amplitude corresponding to the diagram in Fig.~\ref{fig.DD},
\ba
&&	\mathcal{M}_D =
    \frac{g_e}{16 \pi^2}
    \frac{[\bar{v}(p_1) \gamma_\mu u(p_2)]}{q^2-M_\psi^2+i M_\psi\,\Gamma_\psi}
    \cdot
    \int\frac{dk}{i \pi^2}
    \frac{[\bar{u}(q_1) \gamma_5 (\hat{k} + M_\Xi) \gamma_5 v(q_2)] (2 k + q_2 - q_1)^\mu}{(k^2 - M_\Xi^2)((k-q_1)^2 - M_D^2) ((k+q_2)^2 - M_D^2)}
    \times\nn\\
&& \times G_{\psi D\bar{D}}(q^2,(k+q_2)^2,(k-q_1)^2) G_{\Xi D \Sigma}(k^2,(k-q_1)^2) G_{\Xi D \Sigma}(k^2,(k+q_2)^2),
    \label{AmplitudePsiDD}
\ea
where $M_D$, $M_\psi$, and $M_\Xi$ are the masses of the $D$-meson, $\psi (3770)$-charmonium and $\Xi$-hyperon, respectively.
Thus, we use the following dependence of the form factor on the $\psi D\bar{D}$ vertex:
\ba
	G_{\psi D\bar{D}} (s,M_D^2,M_D^2) =
	g_{\psi D\bar{D}} \, \frac{M_\psi^2}{s} \, \frac{\log (M_\psi^2/\Lambda_D^2)}{\log (s/\Lambda_D^2)},
	\label{PsiDDFormfactor}
\ea
where the constant $\Lambda_D$ is fixed on the characteristic value of the reaction $\Lambda_D = 2 M_D$.
Comparing this expression with the general form of the amplitude from \eqref{Mpsi}, we can determine the current
$J^{\psi\to \Sigma^0\bar{\Sigma}^0}_\nu (q)$ in the following form:
\ba
&& J^{\psi\to \Sigma^0\bar{\Sigma}^0}_\mu (q) = \frac{1}{16 \pi^2}
\int\frac{dk}{i \pi^2}
    \frac{[\bar{u}(q_1) \gamma_5 (\hat{k} + M_\Xi) \gamma_5 v(q_2)] (2 k + q_2 - q_1)^\mu}{(k^2 - M_\Xi^2)((k-q_1)^2 - M_D^2) ((k+q_2)^2 - M_D^2)}
   \times\nn\\
&& \times G_{\psi D\bar{D}}(q^2,(k+q_2)^2,(k-q_1)^2) G_{\Xi D \Sigma}(k^2,(k-q_1)^2) G_{\Xi D \Sigma}(k^2,(k+q_2)^2),
\label{JDD}
\ea
and it is possible insert it into \eqref{Si}.
In this case, for $S_D$ \eqref{Si} we obtain the following expression:
\ba
&&	S_D(s) = \frac{\alpha g_e \beta G(s)}{48 \pi^2 s} \int\frac{dk}{i \pi^2}
    \frac{Sp D(s, k^2)}{(k^2 - M_\Xi^2)((k-q_1)^2 - M_D^2) ((k+q_2)^2 - M_D^2)}
   \times\nn\\
&&  \times G_{\psi D\bar{D}}(q^2,(k+q_2)^2,(k-q_1)^2) G_{\Xi D \Sigma}(k^2,(k-q_1)^2) G_{\Xi D \Sigma}(k^2,(k+q_2)^2 = \nn \\
&&  \alpha_{D}(s) Z_D(s).
   \label{SDG}
\ea
Here, $\alpha_{D}(s)$ and  $Z_D(s)$ has the following forms:
\ba
&& \alpha_{D}(s) = \frac{\alpha g_e \beta G(s)}{48 \pi^2}, \nn \\
&& Z_D(s) = \frac{1}{s} \int\frac{dk}{i \pi^2}
    \frac{Sp D(s, k^2)}{(k^2 - M_\Xi^2)((k-q_1)^2 - M_D^2) ((k+q_2)^2 - M_D^2)}
   \times\nn\\
&&  \times G_{\psi D\bar{D}}(q^2,(k+q_2)^2,(k-q_1)^2) G_{\Xi D \Sigma}(k^2,(k-q_1)^2) G_{\Xi D \Sigma}(k^2,(k+q_2)^2.
  \label{ZD}
\ea
where $Sp D(s, k^2)$ is the trace of the Dirac matrices over the baryon line that can be written as follows:
\vspace*{-0.5cm}
\ba
&&	Sp D(s, k^2) =
     Sp [(\hat {q_1}+M_\Sigma) \gamma_5 (\hat {k}+M_\Xi) \gamma_5 (\hat {q_2}-M_\Sigma) (\hat {k} - M_\Sigma)]
    =\nn\\
&&  = 2\left((k^2)^2 + k^2 (s - 2 (M_D^2 + M_\Sigma M_\Xi)) - s M_\Sigma M_\Xi + c_D  \right),
    \label{SpDExplicit}
\ea
where $c_D$ has the following form:
\ba
&c_D
    =
    M_D^4 + 2 M_\Sigma M_\Xi M_D^2 + 2 M_\Xi M_\Sigma^3 - M_\Sigma^4.
    \label{cD}
\ea
In \eqref{SDG}, the quantities $G_{\psi D\bar{D}}$ and $G_{\Xi D \Sigma}$ in \cite{I9,I37} are the form factors
for the vertices $\psi \to D\bar{D}$ and $D \to \Xi \Sigma$. \\
From \eqref{ZD} we can calculate the quantity of $Z_D(s)$ .
By using the Cutkosky rule \cite{Cutkosky} the $D$-meson propagators are equivalently replaced with the delta function,
\ba
  &\frac{1}{(k-q_1)^2 - M_D^2}
	\quad\longrightarrow\quad
	- 2 \pi i ~ \delta ((k-q_1)^2 - M_D^2) ~\theta (-(k-q_1)_0),
   \nn\\
 &\frac{1}{(k+q_2)^2 - M_D^2}
	\quad\longrightarrow\quad
	- 2 \pi i ~ \delta ((k+q_2)^2 - M_D^2) ~\theta ((k+q_2)_0),
\label{delta}
\ea
By using these two $\delta$-functions \eqref{delta} in \eqref{ZD}, we obtain the imaginary part of this quantity from $Z_D(s)$,
\ba
&&	2i \, \mbox{Im} \, Z_D(s)
	=\frac{(-2\pi i)^2}{s}
    \int\frac{dk}{i \pi^2}
    \frac{Sp D(s, k^2)}{k^2 - M_\Xi^2}
    G_{\psi D\bar{D}}(s,(k+q_2)^2,(k-q_1)^2)
    \times\nn\\
&&   \times
    G_{\Xi D \Sigma}(k^2,(k-q_1)^2)~G_{\Xi D \Sigma}(k^2,(k+q_2)^2)
    \delta ((k+q_2)^2 - M_D^2)~\delta ((k-q_1)^2 - M_D^2)
    \times\nn\\
&&    \times
    \theta ((k+q_2)_0)~\theta (-(k-q_1)_0).
    \label{ImZD0}
\ea
After replacing these two $\delta$-functions in \eqref{ImZD0} and performing loop integrations, we obtain the final expression
for the imaginary part of this quantity,
\ba
&&	\mbox{Im} \, Z_D\br{s}=
	-\frac{2\pi}{s^{3/2}}
	G_{\psi D\bar{D}}(s,M_D^2,M_D^2)
	\int\limits_{C_k^{(1)}}^1 \frac{dC_k}{\sqrt{D_1}}
    \sum_{i=1,2}
    \frac{k_{(i)}^2} {k_{(i)}^2 + M_\Xi^2}~
	\times\nn\\
&& \quad\times
    SpD(s,-k_{(i)}^2)~G_{\Xi D \Sigma}^2(-k_{(i)}^2,M_D^2),
    \qquad
	s > 4M_D^2,
    \label{ImZD}
\ea
where $C_k = \cos\theta_k$ is the polar angle.
The definitions of the quantities $k_{(i)}$, $D_1$ and $C_k^{(1)}$  we will be written in the following form:
\ba
	k_{(1,2)} = \frac{1}{2} (\sqrt{s} \, \beta \, C_k \pm \sqrt{ D_1 }),
	\,\,\,
	D_1 = s \, \beta^2 \, C_k^2 - 4 (M_D^2 - M_{\Sigma}^2),
    \,\,\,
    C_k^{(1,2)} = \pm \frac{2}{\beta} \sqrt{\frac{M_D^2 - M_{\Sigma}^2}{s}}.
\ea
To evaluate $\mbox{Im} \, Z_D$ from (\ref{ImZD}), we need to consider an explicit form factor expression.
Thus, for the $\psi \to D\bar{D}$ vertex, we need only the dependence over charmonium virtuality  $q^2 = s$.
It is only in this region that we will be interested in the dependence of the function $G_{\psi D\bar{D}}(s,M_D^2,M_D^2)$ .
We need to note that when computing the imaginary part of the $Z_D$ quantity using expression (\ref{ImZD}),
$D$-meson legs are on mass shell.

We need to note that the details of the calculation of the quantity $Z_D\br{s}$ can be found in \cite{Ahmadov, Bystritskiy}.
Here we technically calculate the imaginary part of this quantity and then restore its real part by using the dispersion
relation with one subtraction at $q^2=0$.
It must be noted that the $\Sigma$-hyperon (which is the $uds$ quark state) does not have open charm and, therefore,
the vertex $\psi \to \Sigma^0 \bar{\Sigma}^0$ at $q^2=0$ is zero.  \\
First, we fix the normalization of this vertex function $G_{\psi D\bar{D}}(s,M_D^2,M_D^2)$ in the decay $\psi(3770)$ into $D\bar{D}$.
For this, it is convenient to use the quantity of the decay width of $\psi \to D \bar{D}$, which fixes the functions,
\vspace*{-0.5cm}
\ba
g_{\psi D\bar{D}} \equiv G_{\psi D\bar{D}}(M_\psi^2, M_D^2, M_D^2).
\ea
Cutting the diagram by $D$-meson propagators, we get the vertex $\psi \to D\bar{D}$ with the only dependence on the
charmonium virtuality $q^2 = s$, whose $D$-meson legs are on mass shell.

To find the value of the constant $g_{\psi D\bar{D}}$, we need to calculate the width of the $\psi \to D \bar{D}$ charmonium decay.
After calculating the decay width with the use of the standard formula, we obtain an expression for the total decay width in the following form:
\ba
\Gamma_{\psi D\bar{D}} = \frac{g_{\psi D\bar{D}}^2 M_{\psi}\beta_{D}^3}{48 \pi}.
\ea
From here, knowing the experimental value for the width of the decay of the charmonium $\Gamma_{\psi D\bar{D}} = 25 \,MeV$ \cite{I16},
we can find the quantity of the constant $g_{\psi D\bar{D}}$ as
\ba
g_{\psi D\bar{D}} = 4 \sqrt{\frac{3 \pi \Gamma_{\psi D\bar{D}}}{M_{\psi}\beta_{D}^3}} \approx 18.4,
\ea
where $\beta_D = \sqrt{1 - 4M_D^2/M_\psi^2}$ is the $D$-meson velocity in this decay.

Now we will consider the function $G_{\Xi D \Sigma} (k^2,p^2)$ from (\ref{ZD}). Once again, the only dependence in the
imaginary part of $Z_D$ is the off-mass shell of the $\Xi$ baryon in the $t$-channel, since $k^2 < 0$.

In \cite{Ahmadov,Bystritskiy,BA}, we used the following form of the $\Lambda D P$-vertex based on the results of \cite{Reinders,Navarra}.
However, in this work we will use a form of the $\Xi D \Sigma$-vertex that corresponds to the results of \cite{Reinders, Choe}.

The $SU(4)$ symmetry leads us to the same result for $G_{\Xi D \Sigma}$,
\ba
	G_{\Xi D \Sigma}(k^2, M_D^2) = \frac{f_D \, g_{\Xi D \Sigma}}{m_u + m_c},
	\qquad
	k^2 < 0,
\ea
where $f_D \approx 180~\MeV$ and
\vspace*{-0.5cm}
\ba
	g_{\Xi D \Sigma} \approx g_{K \Sigma \Xi} = -7.02.
	\label{gLambdaDXi}
\ea
For quark masses, the following values are used: $m_u \approx 280~\MeV$ and  $m_c = 1.27~\GeV$ \cite{I16}.
\section{The three-gluon mechanism}
\label{sec.ThreeGluonMechanism}
\begin{figure}
	\centering
    \includegraphics[width=0.50\textwidth]{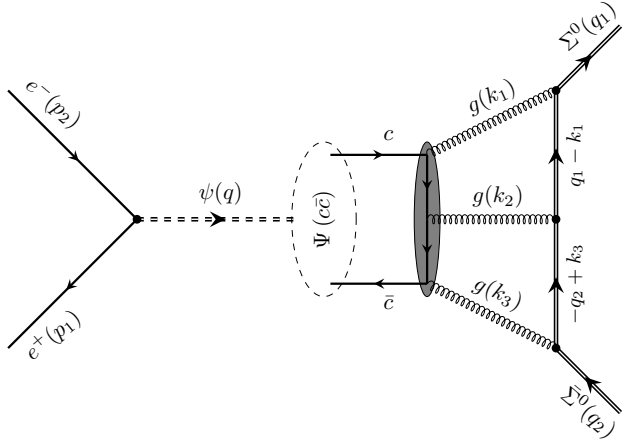}
    \caption{Three-gluon mechanism of the Feynman diagram contributing to the $\Sigma^0\bar{\Sigma}^0$ production
    in the process $e^+e^- \to \Sigma^0\bar{\Sigma}^0$.}
    \label{fig.3G}
\end{figure}

In this section, we will consider the contribution of the intermediate charmonium with the transition to the final $\Sigma^0 \bar{\Sigma}^0$-pair
through the mechanism of three-gluon annihilation.
To calculate the contribution to the cross section, we need to calculate the quantity of $S_{3g}$ from  \eqref{Si}, which is included in
\eqref{SigmaIntViaSi}.
The corresponding Feynman diagrams for the three-gluon mechanism is presented in Fig.~\ref{fig.3G}.
We would like to note that the three-gluon mechanism was considered for the first time in \cite{Ahmadov,Bystritskiy};
it was refined and some typos and minor errors were corrected.
In this paper, we simply apply this mechanism to the production of the $\Sigma^0\bar{\Sigma}^0$-hyperon pair through a three-gluon exchange.

According to the Feynman diagram (Fig.~\ref{fig.3G}),  we write the following contribution to the quantity $S_{3g}$ from \eqref{SigmaIntViaSi}
[which coincide with Eqs. (16) and (17) from \cite{Ahmadov}] to the interference of a charmonium state with the Born amplitude [see (\ref{Si})]:
\ba
    S_{3g}(s) = \alpha_{3g} (s) \, Z_{3g}(s),
    \label{S3g}
\ea
where
\ba
&&    \alpha_{3g} (s) = \frac{\alpha\, \alpha_s^3}{2^3 \, 3} g_e \, g_{col} \, \phi \, \beta \, G (s) \, G_\psi(s),
    \label{alpha3g}
    \\
&&    Z_{3g} (s)= \frac{4}{\pi^5 s}\int \frac{dk_1}{k_1^2}\frac{dk_2}{k_2^2}\frac{dk_3}{k_3^2}
    \frac{Sp3g ~ \delta (q-k_1-k_2-k_3)}{((q_1 - k_1)^2 - M_\Sigma^2) ((q_2 - k_3)^2 - M_\Sigma^2)},
    \label{Z3g}
\ea
where $g_{col}$ is the color factor, $g_{col} (1/4) = \left<\Sigma| d^{ijk} ~ T^i T^j T^k |\Sigma \right> = 15/2$.
In \eqref{Z3g}, $Sp3g$ is the product of traces over the $\Sigma$-hyperon and the $c$-quark lines,
\ba
&&	Sp3g = \Sp [\hat{Q}_{\alpha\beta\gamma} (\hat {k}_c + m_c) \gamma^\mu (\hat {k}_{\bar{c}} - m_c)] \cdot
    \Sp\left[(\hat {q}_1+M_\Sigma) \gamma^\alpha (\hat{q}_1 - \hat {k}_1 + M_\Sigma) \gamma^\beta \right.
	\times     \nn \\
&& \qquad\times \left. (-\hat {q}_2 + \hat{k}_3 + M_\Sigma) \gamma^\gamma (\hat {q}_2-M_\Sigma) \gamma_\mu \right],
   	\nn
\ea
here
\ba
&&    \hat{Q}_{\alpha\beta\gamma}  =
    \frac{\gamma_\gamma (-\hat {k}_{\bar{c}} + \hat {k}_3 + m_c) \gamma_\beta (\hat {k}_c - \hat {k}_1 + m_c) \gamma_\alpha}
    {((k_{\bar{c}}-k_3)^2 - m_c^2) ((k_c-k_1)^2 - m_c^2)} +
     \frac{\gamma_\beta (-\hat {k}_{\bar{c}} + \hat {k}_2 + m_c) \gamma_\gamma (\hat {k}_c - \hat {k}_1 + m_c) \gamma_\alpha}
    {((k_{\bar{c}}-k_2)^2 - m_c^2) ((k_c-k_1)^2 - m_c^2)} +  \nn \\
&& + \frac{\gamma_\gamma (-\hat {k}_{\bar{c}} + \hat {k}_3 + m_c) \gamma_\alpha (\hat {k}_c - \hat {k}_2 + m_c) \gamma_\beta}
    {((k_{\bar{c}}-k_3)^2 - m_c^2) ((k_c-k_2)^2 - m_c^2)} +
    \frac{\gamma_\alpha (-\hat {k}_{\bar{c}} + \hat {k}_1 + m_c) \gamma_\gamma (\hat {k}_c - \hat {k}_2 + m_c) \gamma_\beta}
    {((k_{\bar{c}}-k_1)^2 - m_c^2) ((k_c-k_2)^2 - m_c^2)} +  \nn \\
&& + \frac{\gamma_\beta (-\hat {k}_{\bar{c}} + \hat {k}_2 + m_c) \gamma_\alpha (\hat {k}_c - \hat {k}_3 + m_c) \gamma_\gamma}
    {((k_{\bar{c}}-k_2)^2 - m_c^2) ((k_c-k_3)^2 - m_c^2)} +
    \frac{\gamma_\alpha (-\hat {k}_{\bar{c}} + \hat {k}_1 + m_c) \gamma_\beta (\hat {k}_c - \hat {k}_3 + m_c) \gamma_\gamma}
    {((k_{\bar{c}}-k_1)^2 - m_c^2) ((k_c-k_3)^2 - m_c^2)}. \,\,\,\,\,\,\,\,\,\,\,\,
    \label{DefinitionHatQ}
\ea
In order to correctly normalize the color wave function, we use Eq.(5) in (\cite{Chiang}).
The color wave function normalized to unity should have the form,
\ba
\frac{1}{\sqrt{3}}\bar{q}_i q_i = \frac{1}{\sqrt{3}} (\bar{q}_1 q_1 + \bar{q}_2 q_2 + \bar{q}_3 q_3), \nn
\ea
where the factor $1/\sqrt{3}$ provides the correct normalization of this state by 1.

We would like to note that the parameter of $\phi$ in (\ref{alpha3g}) is related to the charmonium wave function and can be written as
\ba
	\phi = \frac{|\psi (\bf {r}=\bf{0})|}{M_\psi^{3/2}} = \frac{\alpha_s^{3/2}}{3\sqrt{3\pi}},
	\label{eq.phi}
\ea
this quantity is obtained from $\psi \to 3 \,g$ decay rate on mass shell.
From Eqs. (44) and (47) it can be seen that the three-gluon mechanism is very sensitive to this value, since at the charmonium
scale (for $s \sim M_c^2$) it depends to a rather high degree on its value.
We would like to note that in the calculation we use the value $\alpha_s(M_c) = 0.28$, which is expected by the evolution of
$\alpha_s$ in QCD from the $b$-quark scale to the $c$-quark scale.
We should like to note that for the charmonium $J/\psi$, a much smaller value $\alpha_s(M_c) = 0.19$ \cite{Chiang},
was used which differs from our case.

It is to be noted that one of the most important corrections concerns the final $\Sigma^0 \bar {\Sigma}^0$ state.
At the $\psi(3770)$ decays, three gluons are obtained, which produce three quark-antiquark pairs and further they form
$\Sigma^0 \bar {\Sigma}^0$ in the final state.
In order to implement this mechanism, we need to reproduce the absolute value of the cross section.
One of the purposes of this mechanism is the transition of three gluons (with the total angular momentum equal to $1$) into the final
$\Sigma^0 \bar {\Sigma}^0$ pair.

Based on the work of \cite{Bystritskiy,BA}, we assume that this mechanism has much in common with the production of a proton-antiproton and
$\Lambda \bar {\Lambda}$ pairs from a photon in a timelike region.
We accept that the factor $G_\psi(s)$ in (\ref{alpha3g}) is the form factor which describes the mechanism of transition of three gluons
into the final $\Sigma^0 \bar{\Sigma}^0$ pair.
Therefore, we can insert into (\ref{alpha3g}) an additional form factor similar to (\ref{Formfactor}) but with a different value of the $C_\psi$ parameter,
\ba
	|G_\psi(s)| = \frac{C_\psi}{s^2 \log^2 (s/\Lambda^2)}.
	\label{FormfactorPsi}
\ea
In this work, when calculating the constant $C_\psi$, we use the same value as in the case of the production of the proton-antiproton and
the $\Sigma^0 \bar{\Sigma}^0$ pair \cite{Bystritskiy,BA}, since gluons do not feel the flavour of the  quarks in the final baryons,
\ba
	C_\psi = (45 \pm 9)~\GeV^4.
	\label{eq.CPsi}
\ea
Using the technique of the dispersion relation, we restore the real part of $Z_D$ by (\ref{ZD}) and $Z_{3g}$ by (\ref{Z3g}).
All details of these calculations are described in \cite{Ahmadov}.
Thus, for the real part of $Z_D$ and $Z_{3g}$, we obtain the following expression:
\ba
	\mbox{Re} \, Z_i(\beta)	=
	\frac{1}{\pi} \left\{
		\mbox{Im} \, Z_i(\beta) \log \left|\frac{1-\beta^2}{\beta_{\text{min}}^2 - \beta^2}\right|
		+ \int\limits_{\beta_{\text{min}}}^1 \frac{2 \beta_1 d\beta_1}{\beta_1^2 - \beta^2}
		[\mbox{Im} \, Z_i(\beta_1) - \mbox{Im} \, Z_i(\beta)]\right\}.
	\label{DispersionRelation}
\ea
Here, we want to note that the imaginary part $\mbox{Im} \, Z_D(\beta)$ of (\ref{ImZD}) for the $D$-meson loop contribution
is nonzero above the threshold ($s > 4M_D^2$); in this case, the lower integration limit in {(\ref{DispersionRelation})
is $\beta_{\text{min}} = \sqrt{1 - M_{\Sigma}^2/M_D^2}$.
However, the threshold of the imaginary part $\mbox{Im} \, Z_{3g}(\beta)$ for the three-gluon contribution coincides with the
reaction threshold, i.e., $s_{\text{min}} = 4M_{\Sigma}^2$, and therefore, the lower limit of integration is $\beta_{\text{min}} = 0$.

\section{The Numerical results}
\label{sec.Numerical}
In this section, we present numerical results by explicitly considering the distribution of the total cross-section from the
total energy in electron-positron collisions at the BESIII \cite{I37} and BABAR \cite{I9} energies.
The results obtained by us are compared with the experimental data of the BESIII and BABAR.
We calculate the total cross section using formula \eqref{TotalCrossSectionBorn} for $\Sigma^0 \bar{\Sigma}^0$ pair production
processes as a function of the collider c.m. energy $\sqrt s$ in the range from 2.3864 GeV to 4.6 GeV.

In Fig.~\ref{WideRange}, we plot the dependence of the total cross section on c.m. energy $\sqrt s$ for $e^+ e^- \to \Sigma^0 \bar{\Sigma}^0$
in the Born approximation.
It can be seen that with growth in energy, the total cross section increases sharply and in the value of energy $\sqrt {s}$ = 2.4806 \, GeV,
the cross section reaches a maximum, further, with increasing energy, the cross section decreases.
The obtained result on the total cross section in the Born approximation is shown in Fig.~\ref{WideRange}.
The obtained theoretical result is compared with experimental data from the BESIII and BABAR.

We want to note that the main parts for the general cross section are the quantities $Z_D (s)$ from \eqref{ZD} and
$Z_{3g}\br{s}$ from \eqref{Z3g}, which give the corresponding ($D$-meson loop and three gluons) contributions.
The dependence of $Z_D (s)$ on the total energy of $\sqrt{s}$ in the range starting from the reaction threshold
$\sqrt{s}= 2M_\Sigma^0$ to $4.6~$GeV is shown in Fig.~\ref{fig.ZD}.
It is seen that in the real and imaginary parts of $Z_D (s)$ the quantity remains the same as in the case of the
$p \bar {p}$ [Fig.~7(a) in \cite{Bystritskiy}], and $\Lambda \bar {\Lambda}$ [Fig.~6 in \cite{BA}] final states.
In Fig.~\ref{fig.Z3g}, we illustrated the dependence of the real and imaginary parts of $Z_{3g} (s)$ from the total energy $\sqrt{s}$.
It can be seen from this figure that the same general behavior of the curves as in the case of the $p \bar {p}$ [Fig.~7(b) in \cite{Bystritskiy}],
and $\Lambda \bar {\Lambda}$ [Fig.~7 in \cite{BA}] final states, but the difference in the numerical results is much more noticeable.
However, the characteristic large negative values of the quantity $Z_{3g} (s)$ remains, which gives a large relative phase with respect
to the Born contribution to the amplitude.
Figures \ref{fig.ZD} and \ref{fig.Z3g} show that the position of the $\psi(3770)$ resonance is marked with a vertical dashed line.
\begin{figure}
	\vspace{5mm}
	\centering
    \includegraphics[width=0.60\textwidth]{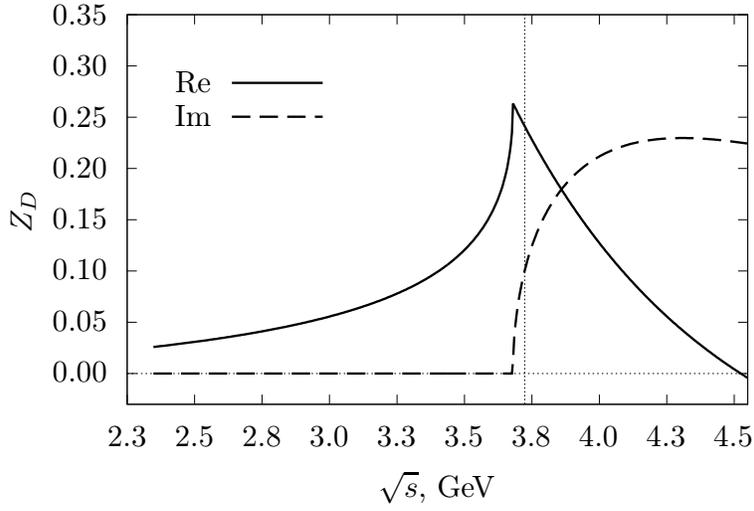}
    \caption{
    	The quantity $Z_D\br{s}$ from (\ref{ZD}) as a function of the c.m. energy $\sqrt{s}$
    starting from the threshold $\sqrt{s} = 2M_\Sigma^0$.
    In this figure the vertical dashed line shows the position of $\psi(3770)$.}
    \label{fig.ZD}
\end{figure}

\begin{figure}
	\vspace{5mm}
	\centering
    \includegraphics[width=0.60\textwidth]{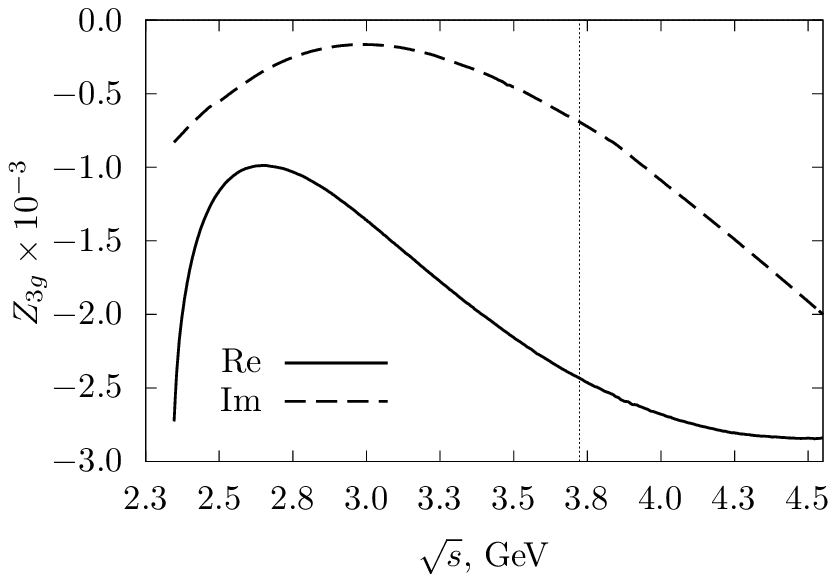}
    \caption{
    	The quantity $Z_{3g}\br{s}$ from (\ref{Z3g}) as a function of the c.m. energy $\sqrt{s}$
     starting from the threshold $\sqrt{s} = 2M_\Sigma^0$.
     In this figure the vertical dashed line shows the position of $\psi(3770)$.}
    \label{fig.Z3g}
\end{figure}
\begin{figure}
	\centering
    \includegraphics[width=0.50\textwidth]{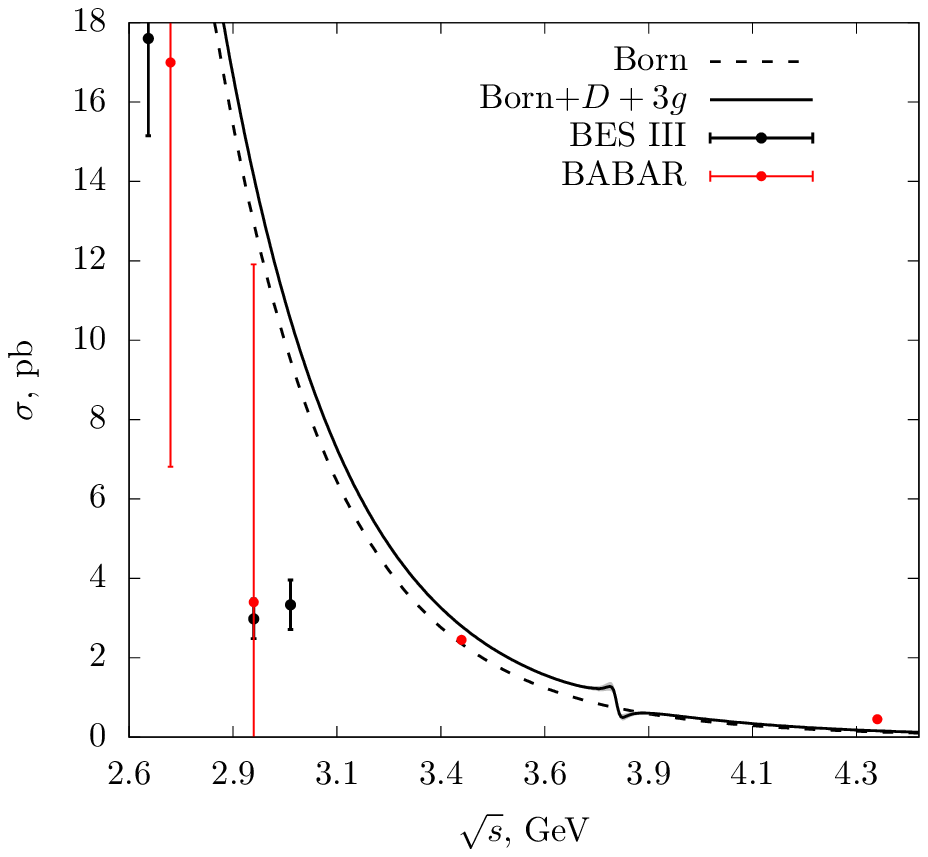}
    \caption{
    	The total cross section in energy region around 2.6 - 4.5 \,GeV including two mechanisms
        ($D$-meson loop  and three gluon) in comparison with the
       BESIII \cite{I37} and BABAR \cite{I9} data.}
    \label{fig.CSNarrow2}
\end{figure}
In Fig.~\ref{fig.CSNarrow2}, we plot the dependence of the total cross section in energy region around 2.6 - 4.5 GeV,
including the $D$-loop and three-gluon mechanisms into the Born cross section, in comparison with the BESIII \cite{I37} and
BABAR \cite{I9} data.
We made this graph, to show the peak near of the $\psi(3770)$ charmonium resonance,
which is now clearly visible.
It should be note that in our model, we consider only the vicinity around $\psi(3770)$.
Thus, we have only this charmonium in the intermediate state.
If one wants to see the plot over a wide interval, then he must include
additional charmonium states, such as $\psi(2S)$ or $\psi(4040)$, for example.
At the moment, we do not want to extend our calculation in such a complicated way.
Therefore, our prediction is valid only in a narrow interval,
say $M(2S)< \sqrt {s}<$ 4040 \,MeV.
We want to note that our models works in the vicinity of charmonium $\psi(3770)$,
so we show only this region.

Since the bin width of the BABAR data is very large, to make a comparison, one should calculate
the following convolution:
\ba
\frac{\int_{x_{min}}^{x_{max}} \sigma(s) dx}{x_{max} - x_{min}},
\label{spm}
\ea
where $x \equiv \sqrt {s}$, \,$x_{\rm {max},\,\, \\ \rm{min}}$ are the limits of bin.
Using the formula \eqref{spm}, we calculated the total cross sections over a wide
range of 3.2 - 3.6 GeV (also in the range 2.8 - 3.0 GeV) and compared our theoretical
results with the experimental BABAR data in the energy ranges of $2.8 - 3.0$ GeV and $3.2 - 3.6$ GeV.
In Table I one can see the corresponding comparison of our results with BABAR data.
\begin{table}
\large{ \begin{tabular}{|c|c|c|c|}
  \hline
    $\sqrt {s},\, GeV$ & $\sigma _{exp} (pb)$   & $\sigma_{th}^{Born+DDloop+ggg} (pb)$    \\
  \hline
    $2.800 - 3.000$ & $3.4_{-7.8 \pm 0.4}^{+8.5}$   & $14.14$    \\
  \hline
    $3.200 - 3.600$ & $<2.5$  & $3.03$  \\
  \hline
  \end{tabular}
  \caption{The numerical results of Born+DD-loop +ggg cross sections for the c.m. energy $\sqrt {s}$ a comparison
  BABAR data.}}
 \label{tab}
\end{table}

In this work, we need to remind that we are not doing any additional parameter fitting.
All parameters of our model are fixed by calculation for the $e^+e^- \to p \bar {p}$ process \cite{Bystritskiy}.

Figure \ref{fig.PhiNarrow} shows the dependence of the total relative phase $\phi_\psi$  as a function of the c.m. energy $\sqrt {s}$,
which is determined by the charmonium contribution of $\mathcal{M}_\psi$ to the amplitude with respect to the Born contribution of $\mathcal{M}_B$
without taking into account the Breit-Wigner factor, that is,
\ba
    S_D (s) + S_{3g} (s) = |S\br{s}| e^{i \phi_\psi},
\ea
\begin{figure}
	\centering
    \includegraphics[width=0.70\textwidth]{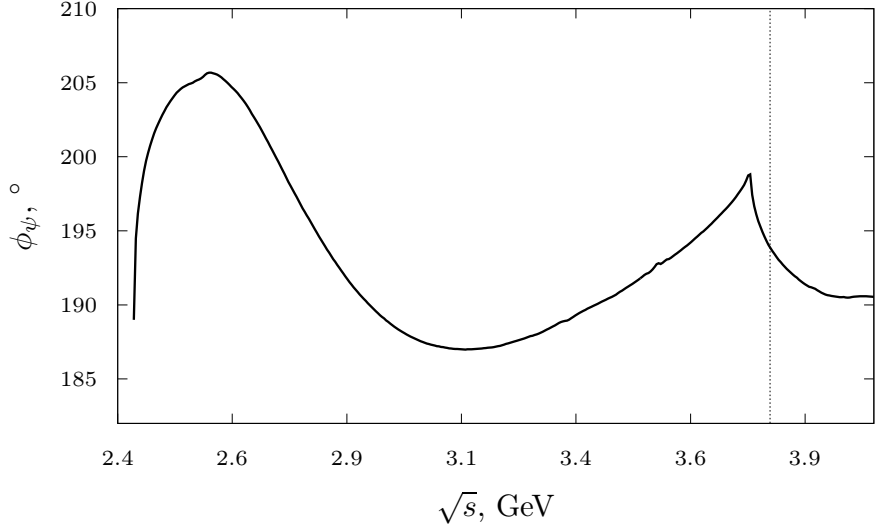}
    \caption{The total relative phase of the charmonium $\psi(3770)$ contribution as a function of the center-of-mass energy $\sqrt{s}$.}
    \label{fig.PhiNarrow}
\end{figure}
where $S_D(s)$ is defined from \eqref{SDG} and $S_{3g}(s)$ from \eqref{S3g}.
It also shows that the position of the $\psi(3770)$ resonance is marked with a vertical dashed line.
It should be noted that, as is seen from Fig.\ref{fig.PhiNarrow} at the point of the $\psi(3770)$ charmonium the relative phase, and the
corresponding total cross section \eqref{TotalCrossSection} at the point of $\psi(3770)$ charmonium are obtained,
\ba
    \sigma_\psi = 1.102~\mbox{pb},
    \qquad
    \phi_\psi = 194^{\circ}.
\ea
It can be assumed that such a feature is common for the decay of charmonium of two baryons into the final state.
This was shown in the $p \bar {p}$ and $\Lambda \bar {\Lambda}$ final states for the charmonium $\psi(3770)$ in \cite{Ahmadov, Bystritskiy,BA} and
for the charmonium $\chi_{c2}(1P)(3556)$ in \cite{Kuraev}.

\section{Conclusion}
\label{sec.Conclusion}
In this paper, we have studied the process of electron-positron annihilation into a $\Sigma^0 \bar{\Sigma}^0$ pair in the
vicinity of the charmonium $\psi(3770)$ resonance.
In the $e^+ e^- \to \Sigma^0 + \bar{\Sigma}^0$ process, besides the Born mechanism, which represents is the pure QED,
we also investigated two more contributions associated with the intermediate state of the charmonium $\psi( 3770)$.
One of them represents the contributions of the $D$-meson loop and the other is the contributions of the three-gluon mechanism.
Here we must remind that the reaction $e^+ e^- \to \Sigma^0 + \bar{\Sigma}^0$ can also be initiated by a vector-charmonium state
such as the $\psi$.
Since the photon and the $\psi$ are both vector mesons, the structures of the corresponding cross section distributions are similar.

It has been shown that both mechanisms make a significant contribution and give a large part of the final result.
It is also important to note that the curve we obtained reproduces a minor (low) slope of the experimental points on the left and right shoulders
with respect to the central point.
Once again, we want to note that in this calculation we do not use any fitting procedures.
All the parameters were fixed for the $p \bar {p}$ production channel in \cite{Bystritskiy}.
We wanted to perform an accurate scan of the energy region around the $\psi(3770)$ charmonium resonance with small steps.
From this we can get a basis for concluding that during the decay of charmonium the phases of the vertices $\psi \to p\bar{p}$, $\psi \to \Lambda\bar{\Lambda}$,
and $\psi \to \Sigma^0 \bar{\Sigma}^0$ are large ($\phi_\psi \sim 200^\circ$) and can be accurately measured in these channels.
It should be noted that the large-phase generation is also shown by us and in other series of papers \cite{Ahmadov, Kuraev, Bystritskiy,BA}.
In the future, we plan to consider other binary processes of formation of final states induced by annihilation of charmonium.

\section{Acknowledgements}
I am grateful to Dr. Yu. M. Bystritskiy for the useful discussions.

\end{document}